\documentclass[%
 reprint,
 superscriptaddress,
 amsmath,amssymb,
 aps,
prl,
]{revtex4-2}

\usepackage{graphicx}
\usepackage{dcolumn}
\usepackage{bm}

\usepackage{amsmath}
\usepackage{amssymb}
\usepackage{booktabs}
\usepackage{hyperref}
\hypersetup{
    colorlinks,
    citecolor=blue,
    linkcolor=blue,
    urlcolor=blue
}

\usepackage[whole]{bxcjkjatype}
\usepackage[normalem]{ulem}
\usepackage{color}
\usepackage{soul}
\definecolor{americanrose}{rgb}{1.0, 0.01, 0.24}
\definecolor{electricpurple}{rgb}{0.75, 0.0, 1.0}
\definecolor{vividgreen}{rgb}{0.0, 0.6, 0.0}

\begin{document}

\preprint{APS/123-QED}

\title{Heisenberg Scaling in Many-Body Kinetic Uncertainty Relation \\ via Quantum Feedback}

\author{Hayato Yunoki}
 \email{yunoki@biom.t.u-tokyo.ac.jp}
\affiliation{Department of Information and Communication Engineering, Graduate School of Information Science and Technology, The University of Tokyo, Tokyo 113-8656, Japan}

\author{Yoshihiko Hasegawa}%
 \email{hasegawa@biom.t.u-tokyo.ac.jp}
\affiliation{Department of Electrical Engineering and Information Systems, Graduate School of Engineering, The University of Tokyo, Tokyo 113-8656, Japan}

\begin{abstract}
Kinetic uncertainty relations impose a fundamental constraint on counting precision by requiring large dynamical activity to suppress fluctuations. For $N$ independent constituents, the activity is additive and scales as $O(N)$, so the corresponding precision bound improves only as $O(1/N)$. A central question is whether many-body quantum systems can instead achieve Heisenberg scaling, with relative variance suppressed as $O(1/N^{2})$. Such scaling could enable substantially enhanced precision in quantum devices, including quantum clocks. Superradiance appears to provide a natural route because its collective emission activity can scale as $O(N^2)$. We show, however, that superradiance alone fails to achieve Heisenberg scaling because this enhanced activity is transient. We then demonstrate that jump-conditioned quantum feedback stabilizes the superradiant high-activity regime and converts this collective enhancement into Heisenberg scaling. We establish this mechanism analytically using a many-body kinetic uncertainty relation and mean-field analysis, and demonstrate through quantum-trajectory simulations that the actual relative counting fluctuation scales as $O(1/N^{2})$.
\end{abstract}

\maketitle


\paragraph{Introduction.---}
Precision is a central figure of merit for quantum devices whose performance is characterized by the stability of their dynamical behavior, such as quantum clocks and quantum batteries \cite{erkerAutonomousQuantumClocks2017, friisPrecisionWorkFluctuations2018, mitchisonQuantumThermalAbsorption2019, antoniomaringuzmanKeyIssuesReview2024, rinaldiMaximumPrecisionCharging2026a}. In stochastic thermodynamics, kinetic uncertainty relations (KURs) have been established as fundamental limits on the precision of general counting observables \cite{garrahanSimpleBoundsFluctuations2017, diterlizziKineticUncertaintyRelation2019}. For a stationary classical Markov jump process, the KUR gives the bound
\begin{equation}
    \frac{\mathrm{Var}[J_\tau]}
    {\langle J_\tau\rangle^2}
    \geq
    \frac{1}{\mathcal A_\tau},
    \label{eq_classical_kur}
\end{equation}
where $J_\tau$ is a counting observable accumulated over an observation time $\tau$, $\mathcal{A}_\tau$ is the dynamical activity defined as the expected total number of jumps during $\tau$, $\mathrm{Var}[\cdot]$ denotes the variance, and $\langle\cdot\rangle$ denotes the expectation value. This relation shows that suppressing the fluctuations of a counting observable requires a correspondingly large dynamical activity. Recent extensions of KURs and related precision bounds to quantum systems have shown that quantum effects such as coherence can enable precision beyond the limits attainable in classical systems \cite{erkerAutonomousQuantumClocks2017, guarnieriThermodynamicsPrecisionQuantum2019, carolloUnravelingLargeDeviation2019, hasegawaQuantumThermodynamicUncertainty2020, hasegawaThermodynamicUncertaintyRelation2021,  millerThermodynamicUncertaintyRelation2021, hasegawaThermodynamicUncertaintyRelation2022, vanvuThermodynamicsPrecisionMarkovian2022, prechRoleQuantumCoherence2025, yunokiQuantumSpeedLimit2025a, ishidaQuantumComputerBasedVerification2025, honmaInformationthermodynamicBoundsPrecision2026}. A fundamental question is then how far such quantum enhancement can improve the precision.

\begin{figure}[t]
\includegraphics[width=\linewidth]{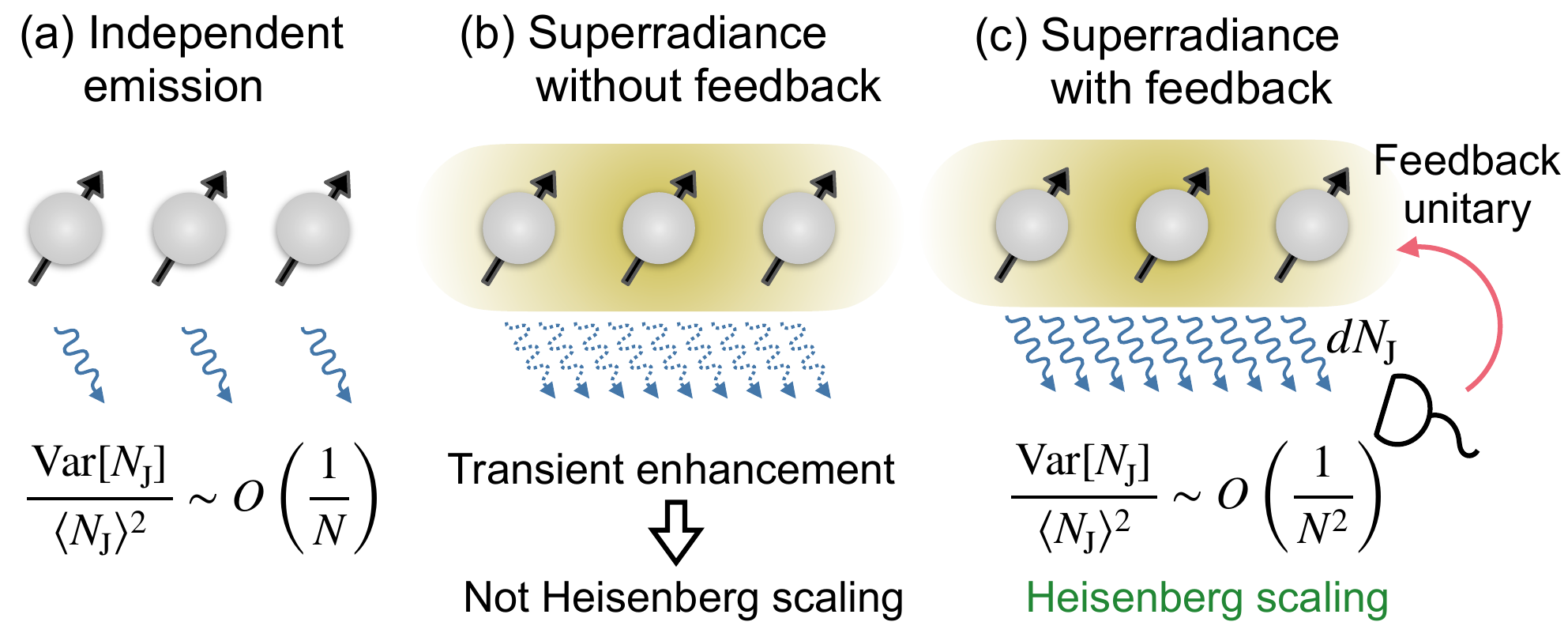}
\caption{\label{fig_MBFBKUR}
Conceptual illustration of Heisenberg-scaling counting precision.
(a) Independent emission yields the standard scaling
$\mathrm{Var}[N_J]/\langle N_J\rangle^2 \sim O(1/N)$.
(b) Superradiance without feedback enhances the emission collectively, but the enhancement is transient and therefore fails to realize Heisenberg scaling.
(c) With jump-conditioned feedback, the superradiant high-activity regime is stabilized, leading to
$\mathrm{Var}[N_J]/\langle N_J\rangle^2 \sim O(1/N^2)$, namely Heisenberg scaling.}
\end{figure}

A prominent example of exploiting quantum effects to enhance performance is quantum metrology \cite{giovannettiAdvancesQuantumMetrology2011}. In parameter estimation with $N$ probes, the estimation variance scales as $O(1/N)$ for classical strategies, whereas by appropriately exploiting quantum correlations, the scaling can be improved to $O(1/N^2)$ \cite{durImprovedQuantumMetrology2014, zhouAchievingHeisenbergLimit2018, koppenhoferDissipativeSuperradiantSpin2022, montenegroQuantumMetrologyBoundary2023, zhouAchievingMetrologicalLimits2024, cabotContinuousSensingParameter2024}. This quadratic improvement is known as Heisenberg scaling. For counting observables constrained by the KUR, $N$ particles contributing independently likewise yield only an $O(1/N)$ scaling of the relative-fluctuation bound, because their dynamical activities add up to $O(N)$, as schematically illustrated in Fig.~\ref{fig_MBFBKUR}(a). This naturally raises the question of whether an analogous Heisenberg scaling can be achieved for the fluctuations of counting observables in open many-body quantum systems. To date, however, the achievability of such $O(1/N^2)$ counting precision has not been established.

Achieving an $O(1/N^2)$ scaling through the KUR requires the dynamical activity accumulated over the observation time to grow as $O(N^2)$. Superradiance appears to offer a natural route \cite{funoSymmetryInducedEnhancement2025, tomaApproachingCarnotEfficiency2026}, since collective spontaneous emission from an ensemble of $N$ identical two-level systems can produce an instantaneous jump rate of $O(N^2)$ \cite{dickeCoherenceSpontaneousRadiation1954a, GROSS1982301}. Our analysis based on the KUR and mean-field equations governing the magnetization dynamics shows, however, that superradiance alone does not achieve Heisenberg scaling of the counting precision. The origin of this failure lies in the transient nature of the collective enhancement [Fig.~\ref{fig_MBFBKUR}(b)]. As emission proceeds, the system is driven away from the high-activity regime, and the initially enhanced jump rate rapidly decays. Consequently, the $O(N^2)$ enhancement cannot be maintained over the observation time required to realize $O(1/N^2)$ counting fluctuations.

To overcome this limitation, we introduce jump-conditioned quantum feedback that stabilizes the system in the high-activity regime [Fig.~\ref{fig_MBFBKUR}(c)]. Jump-conditioned feedback is an established paradigm of quantum control, with applications ranging from quantum state stabilization and quantum error correction to quantum-battery charging \cite{kiesslichReverseQuantumState2012, ahnQuantumErrorCorrection2003, yaoStableChargingRydberg2021}. We analytically show that this feedback sustains the collective enhancement of superradiance, causing the dynamical activity accumulated over the observation time to scale as $O(N^2)$ and thereby making Heisenberg scaling of the counting precision attainable. We further show that the KUR is saturated in the feedback-stabilized regime, implying that the actual relative counting fluctuation itself scales as $O(1/N^2)$, and corroborate this scaling with quantum-trajectory simulations.

\paragraph{Setup.---}
To improve the precision limit imposed by KUR, we consider the Dicke superradiance model as a mechanism for enhancing the dynamical activity. Dicke superradiance is a paradigmatic collective many-body emission process in which $N$ identical two-level systems couple to a common radiation field \cite{dickeCoherenceSpontaneousRadiation1954a,GROSS1982301}. In contrast to independent emission, for which the dynamical
activity grows at most linearly with $N$ (see Supplemental Material \cite{SM_note}), collective emission in the Dicke model can enhance the instantaneous jump rate up to $O(N^2)$. 

The ensemble is described by the collective spin operators
$S_\alpha=\sum_{i=1}^N\sigma_\alpha^{(i)}/2$ for $\alpha=x,y,z$ and $S_\pm=\sum_{i=1}^N\sigma_\pm^{(i)}$, where $\sigma_\alpha^{(i)}$ is the Pauli operator of the $i$th spin and $\sigma_\pm^{(i)}=(\sigma_x^{(i)}\pm i\sigma_y^{(i)})/2$.
The density operator $\rho$ of the ensemble evolves according to the Lindblad equation \cite{goriniCompletelyPositiveDynamical1976a,lindbladGeneratorsQuantumDynamical1976}
\begin{equation}
    \dot{\rho}
    = \mathcal{L}\rho
    =
    -i[\Delta S_z,\rho]
    +
    \gamma
    \left(
        S_-\rho S_+
        -\frac{1}{2}\{S_+S_-,\rho\}
    \right),
    \label{eq_superradiance_master}
\end{equation}
where $\mathcal{L}$ denotes the Liouvillian superoperator, $\Delta$ denotes the transition frequency and $\gamma$ is the collective decay rate, which we hold fixed as $N$ varies
($\gamma=O(1)$).
The second term describes dissipation associated with the collective jump operator
$L=\sqrt{\gamma}S_-$.
For notational simplicity, we henceforth work in the interaction picture with respect to the Hamiltonian $H=\Delta S_z$. In this picture, the Hamiltonian term in Eq.~\eqref{eq_superradiance_master} vanishes, while the dissipative term remains unchanged.
The formulation in the picture with the Hamiltonian is discussed in the Supplemental Material \cite{SM_note}.

To describe the collective dynamics, we introduce the magnetization
$m_\alpha=\langle S_\alpha\rangle/S$.
We consider permutation-symmetric states within the fully symmetric Dicke sector with total spin $S=N/2$.
For sufficiently large $N$, fluctuations of the macroscopic observables $S_\alpha/S$ around their expectation values $m_\alpha$ become negligible, making the mean-field approximation applicable \cite{lanfordObservablesInfinityStates1969, bratteliOperatorAlgebrasQuantum1997, benattiQuantumSpinChain2018, strocchiSymmetryBreaking2021}.
We consider initial states with unit normalized spin length,
$m_x^2+m_y^2+m_z^2=1$, which is conserved by the leading-order
mean-field dynamics for large $N$ \cite{SM_note}.

The Lindblad equation describes the ensemble-averaged dynamics. To characterize the fluctuations of emission events, we unravel the dynamics into quantum-jump trajectories, which corresponds to continuous measurement of the emitted photons \cite{landiCurrentFluctuationsOpen2024}. During an infinitesimal time interval $dt$, the conditional evolution is described by the Kraus operators
\begin{equation}
    M_0 = 1-iH_{\mathrm{eff}}dt,
    \qquad
    M_1 = \sqrt{\gamma dt}\,S_-,
    \label{eq_kraus}
\end{equation}
where $H_{\mathrm{eff}} =-{(i\gamma S_+S_-)}/{2}$.
$M_0$ describes conditional no-jump evolution generated by the non-Hermitian Hamiltonian $H_{\rm eff}$, while $M_1$ describes a quantum jump corresponding to photon emission through the collective jump operator $\sqrt{\gamma}S_-$.

We denote by $dN_{\mathrm J}$ the Poisson increment associated with the detected emission in the time interval $[t,t+dt)$. It takes the value $1$ when a jump occurs and $0$ otherwise. Its conditional expectation is
\begin{equation}
    \mathbb{E}[dN_{\mathrm J}|\rho_c]
    =
    \mathrm{Tr}\!\left[M_1^\dagger M_1\rho_c\right]
    =
    \gamma\langle S_+S_-\rangle_c\,dt,
    \label{eq_jump_rate}
\end{equation}
where $\rho_c$ denotes the conditional state along a quantum trajectory and
$\langle O\rangle_c=\mathrm{Tr}[O\rho_c]$.
The counting observable over an observation time $\tau$ is defined as
\begin{equation}
    N_{\mathrm J}(\tau)
    =
    \int_0^\tau dN_{\mathrm J},
    \label{eq_counting_observable}
\end{equation}
which gives the total number of quantum jumps during the observation time.

\paragraph{KUR for Dicke superradiance.---}
We now derive a KUR for the counting observable $N_\mathrm{J}(\tau)$ in the Dicke superradiance model.
The derivation follows a standard approach widely used for deriving thermodynamic uncertainty relations (TURs) and KURs, in which quantum-jump trajectories are mapped onto a continuous matrix product state \cite{hasegawaQuantumThermodynamicUncertainty2020,hasegawaIrreversibilityLoschmidtEcho2021,hasegawaThermodynamicUncertaintyRelation2022,prechRoleQuantumCoherence2025,honmaInformationthermodynamicBoundsPrecision2026,hasegawaUnifyingSpeedLimit2023,vanvuFundamentalBoundsPrecision2025, yunokiQuantumSpeedLimit2025a}.
Applying the quantum Cram\'er--Rao inequality \cite{hottaQuantumEstimationLocal2004} within this framework, we obtain
\begin{equation}
    \frac{\mathrm{Var}[N_\mathrm{J}(\tau)]}
    {\left[\tau\partial_\tau\langle N_\mathrm{J}(\tau)\rangle\right]^2}
    \geq
    \frac{1}{A_\mathrm{D}(\tau)}.
    \label{eq_mb_kur}
\end{equation}
A detailed derivation of Eq.~\eqref{eq_mb_kur} is given in the End Matter. Here,
\begin{equation}
    A_\mathrm{D}(\tau)
    =
    \gamma\int_0^\tau dt\,\langle S_+S_-\rangle
    \label{eq_activity}
\end{equation}
corresponds to the dynamical activity, defined as the expected total number of quantum jumps during the observation time. In a steady state, $\tau\partial_\tau\langle N_{\mathrm J}(\tau)\rangle=\langle N_{\mathrm J}(\tau)\rangle$, and Eq.~\eqref{eq_mb_kur} reduces to
\begin{equation}
    \frac{\mathrm{Var}[N_{\mathrm J}(\tau)]}
    {\langle N_{\mathrm J}(\tau)\rangle^2}
    \ge
    \frac{1}{A_\mathrm{D}(\tau)}.
    \label{eq_mb_kur_ss}
\end{equation}
Within the mean-field approximation introduced above, this quantity takes the form
\begin{equation}
    A_\mathrm{D}(\tau)\simeq
    \frac{\gamma N^2}{4}
    \int_0^\tau dt\,
    [1-m_z(t)^2].
    \label{eq_activity_mf}
\end{equation}
The derivation of Eq.~\eqref{eq_activity_mf} is given in the Supplemental Material \cite{SM_note}.
Equation~\eqref{eq_activity_mf} shows that, if the magnetization remains in the high-activity region around $m_z=0$, the accumulated dynamical activity can scale as $O(N^2)$, allowing the KUR bounds in Eqs.~\eqref{eq_mb_kur} and \eqref{eq_mb_kur_ss} to exhibit Heisenberg scaling.

Equations~\eqref{eq_mb_kur} and ~\eqref{eq_mb_kur_ss} contain no explicit quantum correction term arising from coherent dynamics, unlike several previously derived quantum KURs \cite{hasegawaQuantumThermodynamicUncertainty2020, nakajimaSymmetriclogarithmicderivativeFisherInformation2023, nishiyamaExactSolutionQuantum2024, prechRoleQuantumCoherence2025, yunokiQuantumSpeedLimit2025a}.
The same form holds in the picture with the Hamiltonian $H=\Delta S_z$ retained, as follows from the coherent--incoherent correspondence established in Ref.~\cite{nishiyamaQuantumThermodynamicUncertainty2025} (see Supplemental Material \cite{SM_note}). Quantum effects nevertheless contribute through the dynamical activity itself, as discussed later.

Although the instantaneous jump rate $\gamma\langle S_+S_-\rangle$ reaches $O(N^2)$ in the superradiant region, this quadratic scaling does not persist in the accumulated activity.
From the Heisenberg-picture equation corresponding to Eq.~\eqref{eq_superradiance_master}, we have
$d\langle S_z\rangle/dt=-\gamma\langle S_+S_-\rangle$,
where we used $\mathcal{L}^\dagger[S_z]=-\gamma S_+S_-$.
Together with Eq.~\eqref{eq_activity}, this gives 
$A_\mathrm{D}(\tau)=\langle S_z(0)\rangle-\langle S_z(\tau)\rangle=O(N)$.
Thus, uncontrolled superradiance cannot provide the $O(N^2)$ accumulated activity required for Heisenberg scaling, despite its $O(N^2)$ instantaneous jump rate.

The origin of this limitation is transparent from the leading-order mean-field dynamics,
\begin{equation}
    \dot m_z
    \simeq
    -\frac{\gamma N}{2}\left(1-m_z^2\right).
    \label{eq_mz_uncontrolled}
\end{equation}
Since $1-m_z^2\geq0$, Eq.~\eqref{eq_mz_uncontrolled} shows that $m_z$ decreases monotonically toward $-1$, driving the system away from the high-activity region around $m_z=0$.
The $O(N^2)$ superradiant jump rate is therefore only transient.
The derivation, including the equations for $m_x$ and $m_y$,
is provided in the Supplemental Material \cite{SM_note}.

\begin{figure}[t]
    \centering
    \includegraphics[width=0.35\textwidth]{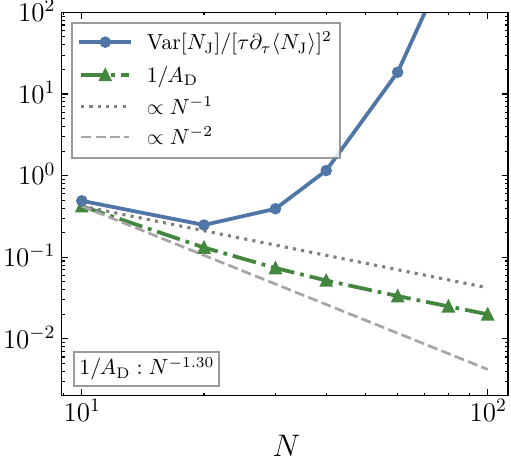}
    \caption{
    \label{fig_no_feedback}
    System-size dependence of the counting fluctuation
    $\mathrm{Var}[N_\mathrm{J}(\tau)]/
    [\tau\partial_\tau\langle N_\mathrm{J}(\tau)\rangle]^2$
    and the inverse dynamical activity $1/A_\mathrm{D}(\tau)$ for uncontrolled Dicke superradiance.
    The initial state is a spin-coherent state with $m_y=1$ and $m_z=0$, and the parameters are $\Delta=0$, $\gamma=0.02$, and $\tau=5.0$.
    The counting fluctuation is evaluated from $10000$ quantum-jump trajectories for
    $N\in\{10,20,30,40,60,80,100\}$.
    The dotted and dashed lines indicate the reference scalings $1/N$ and $1/N^2$, respectively.
    The fitted scaling of $1/A_\mathrm{D}(\tau)$ over the displayed system sizes is also shown.
    }
\end{figure}

We next examine the resulting counting statistics by quantum-jump simulations.
Starting from a spin-coherent state in the high-activity region with $m_y=1$ and $m_z=0$, we generate quantum-jump trajectories of the uncontrolled Dicke dynamics and evaluate the fluctuation of the accumulated jump count $N_\mathrm{J}(\tau)$.
Figure~\ref{fig_no_feedback} shows its dependence on the system size $N$.
The blue curve represents the actual counting fluctuation evaluated from the trajectory ensemble, while the green curve shows the corresponding KUR lower bound $1/A_\mathrm{D}(\tau)$.
The counting fluctuation remains above the KUR bound for all $N$, consistently with Eq.~\eqref{eq_mb_kur}.
More importantly, despite the initially superradiant dynamics, the actual fluctuation does not approach the $O(N^{-2})$ scaling and instead increases for sufficiently large $N$.
The simulation parameters and trajectory sampling are specified in the caption of Fig.~\ref{fig_no_feedback}.

These results demonstrate that superradiance alone is insufficient to realize Heisenberg scaling of the counting precision.
Achieving such scaling therefore requires a mechanism that keeps the system in the high-activity region over the observation time.

\paragraph{Heisenberg scaling via quantum feedback.---}
To stabilize the high-activity region, we introduce the direct jump-feedback scheme formulated by Wiseman and Milburn \cite{wisemanQuantumTheoryContinuous1994,Wiseman_Milburn_2009}.
This is the most basic form of quantum feedback, in which the output of the continuous jump (photon detection) measurement is directly fed back to the system.
We write the measurement current associated with the detected quantum jumps as $I(t)=\nu{dN_\mathrm J}/{dt}$, where $\nu$ sets the weight.
The feedback is implemented through the unitary operation
\begin{equation}
    U_{\mathrm{fb}}(t)
    =
    \exp\!\left[-i I(t)F\,dt\right],
    \label{eq_feedback_unitary_continuous}
\end{equation}
where $F$ is a Hermitian feedback operator.
Each detected quantum jump is immediately followed by the unitary kick $U_{\mathrm{fb}}=e^{-i\nu F}$.
Averaging over the measurement-and-feedback trajectories yields the feedback master equation
\begin{equation}
        \dot{\rho}
        =\mathcal L_{\mathrm{fb}}\rho
        =\gamma\left(
            U_{\mathrm{fb}}S_-\rho S_+U_{\mathrm{fb}}^\dagger
            -\frac{1}{2}\{S_+S_-,\rho\}
        \right),
    \label{eq_feedback_master}
\end{equation}
where $\mathcal L_{\mathrm{fb}}$ denotes the Liouvillian superoperator under the jump feedback.

We take the initial spin-coherent state to satisfy $m_z(0)=0$, with its magnetization pointing in an arbitrary direction in the $xy$ plane,
$\bm{m}(0)=(\cos\phi_0,\sin\phi_0,0)$.
We choose $F$ so that the feedback unitary rotates the collective spin toward the positive $z$ direction,
\begin{equation}
    F
    =
    \sin\phi_0\,S_x
    -
    \cos\phi_0\,S_y.
    \label{eq_feedback_generator}
\end{equation}

We next examine whether the feedback can realize Heisenberg scaling and determine how the feedback strength $\nu$ should be designed.
For the feedback-controlled dynamics, the KUR retains the form of Eq.~\eqref{eq_mb_kur}, as derived in the End Matter. By extending the coherent--incoherent correspondence to time-dependent jump operators, we further show that this form remains valid without an explicit quantum correction term upon returning to the picture in which the Hamiltonian $H=\Delta S_z$ is retained and the feedback operator $F$ consequently becomes time dependent (see Supplemental Material \cite{SM_note}).
If the normalized spin length is conserved at unity
($m_x^2+m_y^2+m_z^2=1$) even in the presence of feedback, as we show below for our feedback design, the leading-order mean-field activity retains the form of Eq.~\eqref{eq_activity_mf}.
The effect of feedback on the activity thus enters through the evolution of the magnetization $m_z(t)$.

We therefore examine how the feedback modifies the dynamics of $m_z$.
Within the mean-field approximation, we obtain
\begin{align}
    \dot m_z
    \simeq{}&
    -\frac{\gamma N}{2}q-\gamma m_z
    \nonumber\\
    &+\frac{\gamma N^2}{4}q
    \left[
        m_{\parallel}\sin\nu
        +m_z(\cos\nu-1)
    \right],
    \label{eq_mz_feedback_general}
\end{align}
where $q=m_x^2+m_y^2$ and
$m_{\parallel}=m_x\cos\phi_0+m_y\sin\phi_0$.
The derivation of Eq.~\eqref{eq_mz_feedback_general}, together with the full mean-field equations, is given in the Supplemental Material \cite{SM_note}.

\begin{figure}[t]
    \centering
    \includegraphics[width=0.35\textwidth]{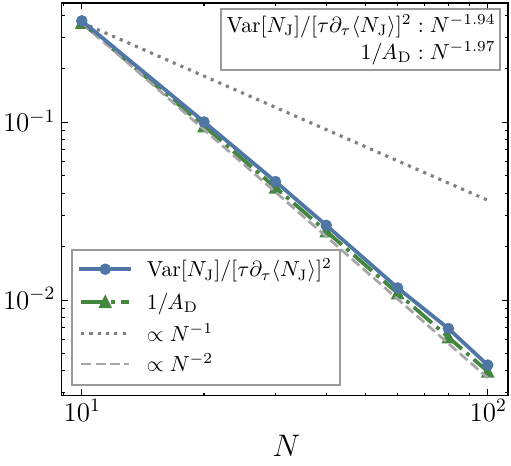}
    \caption{
    \label{fig_feedback_scaling}
    Heisenberg scaling of the counting precision under jump-conditioned feedback.
    The blue curve shows the actual counting fluctuation
    $\mathrm{Var}[N_\mathrm J(\tau)]/
    [\tau\partial_\tau\langle N_\mathrm J(\tau)\rangle]^2$
    evaluated from $10000$ quantum-jump trajectories, and the green curve shows the KUR lower bound $1/A_{\mathrm D}(\tau)$.
    We use the initial spin-coherent state with $\phi_0=\pi/2$ ($m_y=1$ and $m_z=0$), feedback strength $\nu=\arcsin (2/N)$, $\Delta=0$, $\gamma=0.02$, and $\tau=5.0$, for
    $N\in\{10,20,30,40,60,80,100\}$.
    The dotted and dashed lines indicate the reference scalings $1/N$ and $1/N^2$, respectively.
    Power-law fits over the displayed system sizes give exponents $1.94$ for the actual fluctuation and $1.97$ for $1/A_{\mathrm D}(\tau)$.
    }
\end{figure}

For the initial state specified above, we have
$m_z(0)=0$, $q(0)=1$, and $m_{\parallel}(0)=1$.
Equation~\eqref{eq_mz_feedback_general} therefore reduces at $t=0$ to
\begin{equation}
    \dot m_z(0)
    =
    -\frac{\gamma N}{2}
    +
    \frac{\gamma N^2}{4}\sin\nu.
    \label{eq_feedback_balance}
\end{equation}
The leading downward drift is canceled when
$\sin\nu=2/N$, corresponding to $\nu\simeq2/N$ in the large-$N$ regime. For this choice of $\nu$, the mean-field dynamics preserves $m_x^2+m_y^2+m_z^2=1$, as proved in the Supplemental
Material \cite{SM_note}.
With this choice, $m_z=0$ and $m_{\parallel}=1$ form a fixed point of the mean-field dynamics.
At this fixed point, the dynamical activity is maintained as $A_{\mathrm D}(\tau)\simeq{\gamma N^2 \tau}/{4}=O(N^2)$.
The KUR lower bound therefore exhibits Heisenberg scaling under the feedback-controlled dynamics.

Remarkably, the present feedback protocol can also saturate the KUR in the large-$N$ limit.
This implies that the feedback improves not only the KUR lower bound but also the actual counting fluctuation, corresponding to the left-hand side of Eq.~\eqref{eq_mb_kur}, with the same $O(1/N^2)$ scaling. 
The saturation can be understood from the counting statistics.
When the feedback keeps the system at the high-activity state, the conditional jump rate approaches the trajectory-independent constant
$\gamma\langle S_+S_-\rangle_c\simeq\gamma N^2/4$.
The resulting counting process is therefore a homogeneous Poisson process, for which
$\mathrm{Var}[N_\mathrm J(\tau)]=A_{\mathrm D}(\tau)$ and $\langle N_\mathrm J(\tau)\rangle=A_{\mathrm D}(\tau)$.
Since the mean jump count grows linearly with time,
$\tau\partial_\tau\langle N_\mathrm J(\tau)\rangle
=\langle N_\mathrm J(\tau)\rangle$,
we obtain
\begin{equation}
    \frac{\mathrm{Var}[N_\mathrm J(\tau)]}{\langle N_\mathrm J(\tau)\rangle^2}
    =
    \frac{1}{A_{\mathrm D}(\tau)}
    =
    O(1/N^2),
    \label{eq_poisson_saturation}
\end{equation}
showing that the KUR is saturated and that the actual counting fluctuation itself exhibits Heisenberg scaling.

We next test these analytical results by quantum-jump simulations.
For the numerical demonstration, we set $\nu=\arcsin(2/N)$ as obtained above. All other simulation parameters are identical to those used for the uncontrolled dynamics and are given in the caption of Fig.~\ref{fig_feedback_scaling}.
Figure~\ref{fig_feedback_scaling} shows the system-size dependence of the actual counting fluctuation evaluated from the trajectory ensemble and the corresponding KUR lower bound $1/A_{\mathrm D}(\tau)$.
Both quantities closely follow the $O(1/N^2)$ reference scaling, with fitted exponents $1.94$ and $1.97$, respectively.
Thus, the feedback converts the transient superradiant enhancement into an $O(N^2)$ accumulated activity and, more importantly, realizes Heisenberg scaling of the actual counting fluctuation itself. Moreover, the actual fluctuation remains close to $1/A_{\mathrm D}(\tau)$ over the simulated system sizes, indicating near saturation of the KUR.
This behavior is consistent with the Poissonian limit predicted above for the feedback-stabilized dynamics.

The $O(1/N^2)$ scaling raises the question of how quantum coherence contributes to the precision enhancement.
Several quantum KURs contain correction terms arising from coherent dynamics in either the dynamical activity or the terms involving the expectation value \cite{hasegawaQuantumThermodynamicUncertainty2020, nakajimaSymmetriclogarithmicderivativeFisherInformation2023, nishiyamaExactSolutionQuantum2024, prechRoleQuantumCoherence2025, yunokiQuantumSpeedLimit2025a}.
These correction terms can account for precision enhancement beyond classical limits, reflecting the contribution of the smooth evolution described by the Kraus operator $M_0$ to precision.
The KURs derived here contain no such correction terms, either with or without feedback.
Their absence does not imply the absence of quantum effects, which are encoded solely in the jump contribution, $A_{\mathrm D}(\tau)$ itself. The role of quantum coherence can be seen by decomposing the collective jump rate,
\begin{equation}
    \frac{d A_{\mathrm D}(t)}{dt}
=\gamma\langle S_+S_-\rangle
=\gamma\sum_i\langle\sigma^{(i)}_+\sigma^{(i)}_-\rangle
+\gamma \sum_{i\ne j}\langle\sigma^{(i)}_+\sigma^{(j)}_-\rangle.
\end{equation}
The first sum represents emission from individual emitters and is at most $O(N)$.
The second sum arises from interference between emission amplitudes from different emitters, since the collective emission channel does not distinguish which emitter emitted the photon.
Its $N(N-1)$ cross terms can together contribute $O(N^2)$ when their average remains positive and of order unity.
Each cross term $\langle\sigma^{(i)}_+\sigma^{(j)}_-\rangle$ probes coherence between $|e_i g_j\rangle$ and $|g_i e_j\rangle$, since $\sigma^{(i)}_+\sigma^{(j)}_-=|e_i g_j\rangle\langle g_i e_j|$, where $|g_i\rangle$ and $|e_i\rangle$ denote the ground and excited states of spin $i$, respectively.
These states have the same total excitation number, so the relevant coherence lies within degenerate eigenspaces of $H=\Delta S_z$.
This coherence therefore enhances the dynamical activity itself. Feedback sustains this $O(N^2)$ jump rate, yielding an accumulated activity of $O(N^2)$.

The achieved scaling is also optimal within the Dicke superradiance model for counting statistics at fixed $\gamma$ and $\tau$. For any state, $\langle S_+S_-\rangle\leq\lambda_{\max}(S_+S_-)=\|S_+S_-\|$, where $\lambda_{\max}(\cdot)$ denotes the largest eigenvalue and $\|\cdot\|$ the operator norm of $(\cdot)$.
Since $\|S_+S_-\|\leq(N+1)^2/4$, the accumulated activity satisfies
\begin{equation}
A_{\mathrm D}(\tau)
=\gamma\int_0^\tau dt \langle S_+S_-\rangle
\leq \frac{\gamma\tau(N+1)^2}{4}.
\end{equation}
Together with the KUR, this bound implies that the relative counting fluctuation cannot decrease faster than $O(N^{-2})$ under the model.
Feedback enables this optimal scaling to be attained.

\paragraph{Conclusion and outlook.---}
We have shown that Heisenberg scaling of counting precision can be achieved by combining Dicke superradiance with jump-conditioned feedback.
We derived the KUR for the controlled dynamics and used mean-field analysis to determine the feedback strength required to sustain an $O(N^2)$ dynamical activity.
Analysis of the saturation conditions shows how the Poissonian statistics translate this enhanced activity into an actual relative variance of $O(1/N^2)$, as confirmed by quantum-trajectory simulations.
The resulting exponent is optimal within this model for counting statistics.
These results establish feedback as a means of converting transient collective emission into sustained counting precision. Future work could explore analogous improvements in precision scaling can also be achieved in the context of TURs, where precision is constrained by entropy production \cite{baratoThermodynamicUncertaintyRelation2015}.
Feedback-induced precision enhancement in TURs has also been explored \cite{honmaThermodynamicUncertaintyRelation2026a, tojoThermodynamicUncertaintyRelation2026a}.
Furthermore, dynamical activity bounds not only precision but also the speed \cite{shiraishiSpeedLimitClassical2018a, funoSpeedLimitOpen2019, hasegawaUnifyingSpeedLimit2023}.
This motivates investigating how the improved scaling of dynamical activity affects the achievable speed. Our findings offer a new route to qualitative improvements in precision and provide design principles for high-precision quantum devices.


\paragraph{Acknowledgments.---}
This work was supported by JSPS KAKENHI Grant No. JP26K02998, and JST SPRING, Grant Number JPMJSP2108. The quantum-trajectory simulations were performed using QuTiP \cite{johansson2012qutip, johansson2013qutip, qutip5}. The authors used OpenAI ChatGPT (GPT-5.6 Sol) under their direction to assist with simulation code generation, mathematical checks, and manuscript organization. All resulting code, calculations, and text were reviewed and verified by the authors, who take full responsibility for the work.


%

\setcounter{equation}{0}
\renewcommand{\theequation}{A\arabic{equation}}
\renewcommand{\theHequation}{appendix.\arabic{equation}}
\onecolumngrid
\section*{End Matter}
\twocolumngrid
\paragraph*{Appendix: Derivation of the KUR.---}
We derive the KUR in Eq.~\eqref{eq_mb_kur} for the jump-feedback dynamics in Eq.~\eqref{eq_feedback_master}, following the continuous matrix product state (cMPS) approach \cite{hasegawaQuantumThermodynamicUncertainty2020,hasegawaIrreversibilityLoschmidtEcho2021,hasegawaThermodynamicUncertaintyRelation2022, nakajimaSymmetriclogarithmicderivativeFisherInformation2023, nishiyamaExactSolutionQuantum2024, prechRoleQuantumCoherence2025,honmaInformationthermodynamicBoundsPrecision2026,hasegawaUnifyingSpeedLimit2023,vanvuFundamentalBoundsPrecision2025, yunokiQuantumSpeedLimit2025a}. The case without feedback follows from the same derivation by setting the feedback weight $\nu=0$, for which the feedback unitary becomes the identity, $U_\mathrm{fb}=I$.
Here, we work in the interaction picture with $H=0$. The same form of the KUR also holds upon returning to the picture with $H=\Delta S_z$, where the feedback operator becomes time dependent; this case is discussed in the Supplemental Material~\cite{SM_note}.

Let $L=\sqrt{\gamma}S_-$ and $U_\mathrm{fb}=e^{-i\nu F}$. For an infinitesimal time interval $dt$, the measurement operators are described by $M_0=I-iH_{\mathrm{eff}}dt$ and $M_1=\sqrt{dt}\,L$, where $H_{\mathrm{eff}}=-iL^\dagger L/2$. The feedback unitary is applied after a detected jump, so that $U_0=I$ and $U_1=U_\mathrm{fb}$. We introduce the parametrization
\begin{equation}
    L(\theta)=\sqrt{1+\theta}\,L,
    \qquad F(\theta)=F,
    \label{eq_theta_deformation_endmatter}
\end{equation}
while keeping the initial state fixed. This multiplies the right-hand side of Eq.~\eqref{eq_feedback_master} by $1+\theta$, thereby rescaling time by the same factor. For an initial pure state $|\psi_0\rangle$, the joint state of the system and the measurement record is represented by the matrix product state (MPS)
\begin{equation}
    |\Psi_\theta(\tau)\rangle
    =\sum_{\boldsymbol z}U_{z_K}(\theta)M_{z_K}(\theta)\cdots
    U_{z_1}(\theta)M_{z_1}(\theta)|\psi_0\rangle\otimes|\boldsymbol z\rangle,
    \label{eq_discrete_mps_endmatter}
\end{equation}
where $\boldsymbol z=(z_1,\ldots,z_K)$, $z_k\in\{0,1\}$, and $Kdt=\tau$. The states $|\boldsymbol z\rangle$ form an orthonormal basis for the measurement record. Taking the continuous limit $dt\to0$ gives the cMPS. We denote the quantum Fisher information of the cMPS with respect to $\theta$ by $\mathcal J(\theta)$. Considering the estimation of $\theta$ from the counting observable $N_{\mathrm J}(\tau)$, the quantum Cram\'er--Rao inequality gives \cite{hottaQuantumEstimationLocal2004}
\begin{equation}
    \frac{\mathrm{Var}_\theta[N_{\mathrm J}(\tau)]}
    {[\partial_\theta\langle N_{\mathrm J}(\tau)\rangle_\theta]^2}
    \geq
    \frac{1}{\mathcal J(\theta)},
    \label{eq_qcr_endmatter}
\end{equation}
where $\langle\cdot\rangle_\theta$ and $\mathrm{Var}_\theta[\cdot]$ denote the expectation value and variance evaluated for the dynamics parametrized by $\theta$, respectively.
Evaluating this inequality at $\theta=0$, which returns the original dynamics, and using the time rescaling by Eq.~\eqref{eq_theta_deformation_endmatter},
\begin{equation}
    \left.\partial_\theta\langle N_{\mathrm J}(\tau)\rangle_\theta\right|_{\theta=0}
    =\tau\partial_\tau\langle N_{\mathrm J}(\tau)\rangle,
    \label{eq_count_response_endmatter}
\end{equation}
we obtain
\begin{equation}
    \frac{\mathrm{Var}[N_{\mathrm J}(\tau)]}
    {[\tau\partial_\tau\langle N_{\mathrm J}(\tau)\rangle]^2}
    \geq
    \frac{1}{\mathcal J(0)}.
    \label{eq_kur_qfi_endmatter}
\end{equation}

We next evaluate $\mathcal J(0)$. We define $\rho^{\theta,\varphi}(t)
    =\operatorname{Tr}_{\mathrm{record}}
    \left[|\Psi_\theta(t)\rangle\langle\Psi_\varphi(t)|\right]$,
where $\operatorname{Tr}_{\mathrm{record}}$ denotes the partial trace over the measurement-record Hilbert space spanned by $\{|\boldsymbol z\rangle\}$. We also define
\begin{equation}
    C(\theta,\varphi)\equiv\operatorname{Tr}[\rho^{\theta,\varphi}(\tau)]
    =\langle\Psi_\varphi(\tau)|\Psi_\theta(\tau)\rangle.
    \label{eq_overlap_endmatter}
\end{equation}
Then the quantum Fisher information is written as
\begin{equation}
    \mathcal J(0)=4\left[\partial_\theta\partial_\varphi C
    -(\partial_\theta C)(\partial_\varphi C)\right]_{\theta=\varphi=0}.
    \label{eq_qfi_endmatter}
\end{equation}
The system operator $\rho^{\theta,\varphi}$ evolves according to the two-sided master equation
\begin{align}
    \partial_t\rho^{\theta,\varphi}
    ={}&\mathcal L_{\mathrm{fb}}^{\theta,\varphi}\rho^{\theta,\varphi}
    \nonumber\\
    ={}&\sqrt{(1+\theta)(1+\varphi)}\,U_\mathrm{fb}L\rho^{\theta,\varphi}L^\dagger U_\mathrm{fb}^\dagger
    \nonumber\\
    &-\frac{1+\theta}{2}L^\dagger L\rho^{\theta,\varphi}
    -\frac{1+\varphi}{2}\rho^{\theta,\varphi}L^\dagger L,
    \label{eq_twosided_endmatter}
\end{align}
with the parameter-independent initial state of the system $\rho^{\theta,\varphi}(0)=|\psi_0\rangle\langle\psi_0|$. At $\theta=\varphi=0$, $\mathcal L_{\mathrm{fb}}^{\theta,\varphi}$ reduces to the original Liouvillian superoperator $\mathcal L_{\mathrm{fb}}$ in Eq.~\eqref{eq_feedback_master}, and $\rho^{\theta,\varphi}(t)=\rho(t)$.

We define $\mathcal K_1$ and $\mathcal K_2$ by differentiating $\mathcal L_{\mathrm{fb}}^{\theta,\varphi}$ with respect to $\theta$ and $\varphi$, respectively, at $\theta=\varphi=0$. For an arbitrary system operator $X$, their actions are
\begin{align}
    \mathcal K_1X
    &\equiv
    \left.
    \frac{\partial\mathcal L_{\mathrm{fb}}^{\theta,\varphi}}
    {\partial\theta}
    \right|_{\theta=\varphi=0}X=-iH_{\mathrm{eff}}X+\frac12U_\mathrm{fb}LXL^\dagger U_\mathrm{fb}^\dagger,
    \nonumber\\
    \mathcal K_2X
    &\equiv
    \left.
    \frac{\partial\mathcal L_{\mathrm{fb}}^{\theta,\varphi}}
    {\partial\varphi}
    \right|_{\theta=\varphi=0}X=iXH_{\mathrm{eff}}^\dagger+\frac12U_\mathrm{fb}LXL^\dagger U_\mathrm{fb}^\dagger.
    \label{eq_KJ_endmatter}
\end{align}
Since the initial state $\rho(0)=|\psi_0\rangle\langle\psi_0|$ is independent of $\theta$ and $\varphi$, we can write
\begin{equation}
    C(\theta,\varphi)
    =\operatorname{Tr}\!\left[
        e^{\mathcal L_{\mathrm{fb}}^{\theta,\varphi}\tau}\rho(0)
    \right].
\end{equation}
Differentiating the exponential with respect to $\theta$ and evaluating at $\theta=\varphi=0$ gives
\begin{align}
    \left.\partial_\theta C\right|_{\theta=\varphi=0}
    =\int_0^\tau ds\,\operatorname{Tr}\!\left[
        e^{\mathcal L_{\mathrm{fb}}(\tau-s)}
        \mathcal K_1
        e^{\mathcal L_{\mathrm{fb}}s}\rho(0)
    \right].
\end{align}
Here, $e^{\mathcal L_{\mathrm{fb}}s}\rho(0)=\rho(s)$.
Moreover, the physical evolution preserves the trace:
$\operatorname{Tr}[e^{\mathcal L_{\mathrm{fb}}t}X]
=\operatorname{Tr}[X]$ for any system operator $X$.
Using this identity with $X=\mathcal K_1\rho(s)$, and similarly for the derivative with respect to $\varphi$, we obtain
\begin{align}
    \left.\partial_\theta C\right|_{\theta=\varphi=0}
    &=\int_0^\tau ds\,\operatorname{Tr}[
        \mathcal K_1\rho(s)],
    \nonumber\\
    \left.\partial_\varphi C\right|_{\theta=\varphi=0}
    &=\int_0^\tau ds\,\operatorname{Tr}[
        \mathcal K_2\rho(s)].
    \label{eq_overlap_first_derivatives_endmatter}
\end{align}
The cyclic property of the trace and $U_\mathrm{fb}^\dagger U_\mathrm{fb}=I$ give, for any operator $X$,
\begin{equation}
    \operatorname{Tr}[\mathcal K_1X]
    =\operatorname{Tr}[\mathcal K_2X]=0.
    \label{eq_KJ_trace_endmatter}
\end{equation}
Thus both first derivatives in Eq.~\eqref{eq_overlap_first_derivatives_endmatter} vanish.

We next evaluate $\partial_\theta\partial_\varphi C$.
Using trace preservation of the physical evolution, we obtain
\begin{align}
    &\left.\partial_\theta\partial_\varphi C
    \right|_{\theta=\varphi=0}
    \nonumber\\
    &=
    \int_0^\tau ds\,\operatorname{Tr}\!\left[
        \left.
        \partial_\theta\partial_\varphi
        \mathcal L_{\mathrm{fb}}^{\theta,\varphi}
        \right|_{\theta=\varphi=0}\rho(s)
    \right]
    \nonumber\\
    &\quad+
    \int_0^\tau ds_1\int_0^{s_1}ds_2\,
    \operatorname{Tr}\!\left[
        \mathcal K_2
        e^{\mathcal L_{\mathrm{fb}}(s_1-s_2)}
        \mathcal K_1\rho(s_2)
    \right]
    \nonumber\\
    &\quad+
    \int_0^\tau ds_1\int_0^{s_1}ds_2\,
    \operatorname{Tr}\!\left[
        \mathcal K_1
        e^{\mathcal L_{\mathrm{fb}}(s_1-s_2)}
        \mathcal K_2\rho(s_2)
    \right].
\end{align}
From Eq.~\eqref{eq_twosided_endmatter},
\begin{equation}
    \left.
    \partial_\theta\partial_\varphi
    \mathcal L_{\mathrm{fb}}^{\theta,\varphi}X
    \right|_{\theta=\varphi=0}
    =\frac14U_\mathrm{fb}LX L^\dagger U_\mathrm{fb}^\dagger.
\end{equation}
Consequently, the single-integral term is
\begin{align}
    \frac14\int_0^\tau ds\,
    \operatorname{Tr}[U_\mathrm{fb}L\rho(s)L^\dagger U_\mathrm{fb}^\dagger]
    &=\frac14\int_0^\tau ds\,
    \operatorname{Tr}[L^\dagger L\rho(s)]
    \nonumber\\
    &=\frac14A_{\mathrm D}(\tau).
\end{align}
Both double-integral terms vanish by
Eq.~\eqref{eq_KJ_trace_endmatter}, since their integrands
are traces of $\mathcal K_1$ or
$\mathcal K_2$ acting on a system operator.
Therefore,
\begin{equation}
    \left.\partial_\theta\partial_\varphi C
    \right|_{\theta=\varphi=0}
    =\frac14A_{\mathrm D}(\tau).
    \label{eq_overlap_second_derivative_endmatter}
\end{equation}

Together with the vanishing first derivatives, this result reduces Eq.~\eqref{eq_qfi_endmatter} to
\begin{equation}
    \mathcal J(0)=A_{\mathrm D}(\tau)
    =\gamma\int_0^\tau dt\,\langle S_+S_-\rangle.
    \label{eq_qfi_activity_endmatter}
\end{equation}
Substituting Eq.~\eqref{eq_qfi_activity_endmatter} into Eq.~\eqref{eq_kur_qfi_endmatter} establishes the KUR in Eq.~\eqref{eq_mb_kur}.

\end{document}


\setcounter{equation}{0}
\setcounter{figure}{0}
\setcounter{table}{0}
\setcounter{page}{1}
\makeatletter
\renewcommand{\theequation}{S\arabic{equation}}
\renewcommand{\thefigure}{S\arabic{figure}}
\renewcommand{\thetable}{S\arabic{table}}
\renewcommand{\thesection}{S\arabic{section}}
\makeatother

\preprint{APS/123-QED}

\title{Supplemental Material for \\``Heisenberg Scaling in Many-Body Kinetic Uncertainty Relation \\ via Quantum Feedback''}

\author{Hayato Yunoki}
 \email{yunoki@biom.t.u-tokyo.ac.jp}
\affiliation{Department of Information and Communication Engineering, Graduate School of Information Science and Technology, The University of Tokyo, Tokyo 113-8656, Japan}

\author{Yoshihiko Hasegawa}%
 \email{hasegawa@biom.t.u-tokyo.ac.jp}
\affiliation{Department of Electrical Engineering and Information Systems, Graduate School of Engineering, The University of Tokyo, Tokyo 113-8656, Japan}

\newcommand{\refMainKUR}{(6) }
\newcommand{\refMainMFEnoFB}{(10) }
\newcommand{\refMainMFEwithFB}{(14) }
\newcommand{\refMainMFDA}{(9) }

\maketitle

This Supplemental Material provides detailed derivations and additional numerical analyses supporting the results presented in the main text. Equations and figures in this document are numbered with the prefix S (e.g., Eq. (S1) or Fig. S1), while references without this prefix (e.g., Eq. (1) or Fig. 1) refer to the main text.

\section{Kinetic uncertainty relation in the picture retaining the Hamiltonian}
\label{sec:hamiltonian_picture_kur}

The main text uses the interaction picture with respect to $H=\Delta S_z$, in which the Hamiltonian vanishes. Here we retain $H$ explicitly and establish that the counting statistics and the KUR are unchanged. Typically, KURs for dynamics with a nonzero Hamiltonian contain quantum correction terms \cite{hasegawaQuantumThermodynamicUncertainty2020, nakajimaSymmetriclogarithmicderivativeFisherInformation2023, nishiyamaExactSolutionQuantum2024, prechRoleQuantumCoherence2025, yunokiQuantumSpeedLimit2025a}. Under certain conditions, however, Ref.~\cite{nishiyamaQuantumThermodynamicUncertainty2025} establishes a coherent--incoherent correspondence (CIC), whereby the jump statistics are identical to those of the system with the Hamiltonian term removed. This correspondence allows a KUR without quantum correction terms to be obtained even when $H$ is retained. We first apply this result to the dynamics without feedback. We then extend the CIC condition to time-dependent jump operators and show that it applies to our feedback-controlled dynamics, yielding the same KUR as in the main text.

We write
\begin{equation}
    V(t)=e^{-iHt},\qquad H=\Delta S_z,\qquad L=\sqrt{\gamma}S_-,\qquad
    \rho_{\mathrm S}(t)=V(t)\rho_{\mathrm I}(t)V^\dagger(t).
    \label{eq_cic_picture}
\end{equation}
Throughout this section, the subscripts and superscripts
$\mathrm S$ and $\mathrm I$ indicate the picture with
$H$ and the interaction picture, respectively. Both descriptions start from the same density operator $\rho_0$, and $N_{\mathrm J}(\tau)$ counts the detected emission events.

\subsection{KUR without feedback: coherent corrections and the CIC}
\label{subsec:cic_no_feedback}
\label{subsec:cic_corrections}

Without feedback, the Lindblad equation is
\begin{equation}
    \dot\rho_{\mathrm S}
    =\mathcal L\rho_{\mathrm S}
    =-i[H,\rho_{\mathrm S}]+\mathcal D[L]\rho_{\mathrm S},
    \qquad
    \mathcal D[L]X=LXL^\dagger-\frac12\{L^\dagger L,X\}, 
    \label{eq_cic_no_feedback_master}
\end{equation}
where $X$ is an arbitrary system operator. When $H$ is retained, quantum KURs typically contain explicit coherent corrections. For example, Refs.~\cite{hasegawaQuantumThermodynamicUncertainty2020,
nishiyamaExactSolutionQuantum2024} derive the following bound:
\begin{equation}
    \frac{\operatorname{Var}[N_{\mathrm J}(\tau)]}
    {[\tau\partial_\tau\langle N_{\mathrm J}(\tau)\rangle]^2}
    \geq\frac{1}{\mathcal B(\tau)},
    \qquad \mathcal B(\tau)=A_{\mathrm D}(\tau)+\mathcal C(\tau).
    \label{eq_cic_qda_bound}
\end{equation}
Here, the correction term $\mathcal C(\tau)$ is given by
\begin{equation}
    \mathcal C(\tau)
    =8\int_0^\tau ds_1\int_0^{s_1}ds_2\,
    \operatorname{Re}\operatorname{Tr}\!\left[
        H_{\mathrm{eff}}^\dagger\check H(s_1-s_2)\rho_{\mathrm S}(s_2)
    \right]
    -4\left(\int_0^\tau ds\,\operatorname{Tr}[H\rho_{\mathrm S}(s)]\right)^2,
    \label{eq_cic_coherent_correction}
\end{equation}
where $H_{\mathrm{eff}}=H-iL^\dagger L/2$, $\check H(t)=e^{\mathcal L^\dagger t}H$, and $\mathcal L^\dagger$ is the adjoint of $\mathcal L$.

In another approach, a quantum correction enters the response of the mean count rather than the dynamical activity. In the steady state, Ref.~\cite{prechRoleQuantumCoherence2025} gives
\begin{equation}
    \frac{\operatorname{Var}[N_{\mathrm J}(\tau)]}
    {\langle N_{\mathrm J}(\tau)\rangle^2}
    \geq
    \frac{(1+\psi)^2}{A_{\mathrm D}(\tau)},
    \qquad
    \psi=
    \frac{\operatorname{Tr}[
        \mathcal J_{\mathrm c}\mathcal L^D
        \mathcal H\rho_{\mathrm{ss}}]}{j},
    \label{eq_cic_response_correction}
\end{equation}
where $j=\langle N_{\mathrm J}(\tau)\rangle/\tau$,
$\mathcal H X=-i[H,X]$, $\mathcal J_{\mathrm c}X=L XL^\dagger$, and $\mathcal L^D$ is the Drazin inverse of the Liouvillian. Here, $\psi$ represents the quantum correction.

The CIC in Ref.~\cite{nishiyamaQuantumThermodynamicUncertainty2025}
yields a KUR without either type of correction when the
time-independent jump operator satisfies $[L,H]=\omega L$
with real $\omega$. Under this condition, removing the Hamiltonian while keeping the jump operators and the initial state fixed preserves the jump statistics and the dynamical activity. The dynamics without the Hamiltonian term is referred to as incoherent.

We now examine this mechanism in our Dicke superradiance model,
for which $[L_\mathrm S,H]=\Delta L$ and $[H,L_\mathrm S^\dagger L_\mathrm S]=0$. The jump operator in the interaction
picture is therefore
\begin{equation}
    L_{\mathrm I}(t)
    \equiv V^\dagger(t)L_\mathrm S V(t)
    =e^{-i\Delta t}L_\mathrm S.
    \label{eq_cic_jump_interaction}
\end{equation}
Since this phase cancels in the dissipator,
$\mathcal D[L_{\mathrm I}(t)]=\mathcal D[L_\mathrm S]$, the
interaction-picture density operator obeys
\begin{equation}
    \dot\rho_{\mathrm I}(t)=\mathcal D[L_\mathrm S]\rho_{\mathrm I}(t).
    \label{eq_cic_no_feedback_interaction}
\end{equation}
Thus, its evolution is identical to that of the system
with $H=0$.

The phase in $L_{\mathrm I}(t)$ does not affect the
probability of a jump record
$\Gamma=(t_1,\ldots,t_n)$, where $0<t_1<\cdots<t_n<\tau$ are the detected jump times during the observation interval $[0,\tau]$. The CIC therefore gives
\begin{equation}
    P_{\mathrm S}[\Gamma]=P_{\mathrm I}[\Gamma],
    \label{eq_cic_no_feedback_probability}
\end{equation}
so the mean and variance of $N_{\mathrm J}(\tau)$
coincide in the two pictures. 
The dynamical activity is also unchanged:
\begin{equation}
    A_{\mathrm D}^{\mathrm S}(\tau)
    =\int_0^\tau dt\,
      \operatorname{Tr}[L_\mathrm S^\dagger L_\mathrm S\rho_{\mathrm S}(t)]
    =\int_0^\tau dt\,
      \operatorname{Tr}[L_{\mathrm I}^\dagger L_{\mathrm I}\rho_{\mathrm I}(t)]
    =A_{\mathrm D}^{\mathrm I}(\tau),
    \label{eq_cic_activity_no_feedback}
\end{equation}
where we used $\rho_{\mathrm S}(t)
=V(t)\rho_{\mathrm I}(t)V^\dagger(t)$ and
$[H,L_\mathrm S^\dagger L_\mathrm S]=0$.

Applying the $H=0$ KUR
derived in the End Matter therefore yields
\begin{equation}
    \frac{\operatorname{Var}_{\mathrm S}[N_{\mathrm J}(\tau)]}
    {[\tau\partial_\tau
      \langle N_{\mathrm J}(\tau)\rangle_{\mathrm S}]^2}
    \geq\frac{1}{A_{\mathrm D}^{\mathrm S}(\tau)}.
    \label{eq_cic_kur_no_feedback}
\end{equation}
This is the same KUR as Eq.~\refMainKUR of the main
text, even though the Hamiltonian is retained.
The absence of a quantum correction in this bound follows
from the equality of the counting statistics and activity
with those of the $H=0$ system; it does not require either $\mathcal C(\tau)$ in
Eq.~\eqref{eq_cic_coherent_correction} or $\psi$ in
Eq.~\eqref{eq_cic_response_correction}.

\subsection{Feedback operator in the picture retaining the Hamiltonian}
\label{subsec:cic_feedback_picture}
So far, we have considered the dynamics without feedback.
We now turn to the dynamics under jump feedback.
In the interaction picture, the feedback operator of the main text is
\begin{equation}
    F_{\mathrm I}=\sin\phi_0\,S_x-\cos\phi_0\,S_y,
    \qquad U_{\mathrm{fb}}^{\mathrm I}=e^{-i\nu F_{\mathrm I}},
    \qquad J_{\mathrm I}=U_{\mathrm{fb}}^{\mathrm I}L.
    \label{eq_cic_feedback_interaction}
\end{equation}
Here $J_{\mathrm I}$ denotes the combined jump-and-feedback operator. Transforming the same physical protocol to the picture with $H$ gives
\begin{align}
    F_{\mathrm S}(t)
    &=V(t)F_{\mathrm I}V^\dagger(t)
    =\sin(\phi_0+\Delta t)S_x-\cos(\phi_0+\Delta t)S_y,
    \nonumber\\
    U_{\mathrm{fb}}^{\mathrm S}(t)&=e^{-i\nu F_{\mathrm S}(t)}
    =V(t)U_{\mathrm{fb}}^{\mathrm I}V^\dagger(t),
    \qquad J_{\mathrm S}(t)=U_{\mathrm{fb}}^{\mathrm S}(t)L.
    \label{eq_cic_rotating_feedback}
\end{align}
For the choice $\phi_0=\pi/2$ used in the numerical examples, $F_{\mathrm I}=S_x$ and this becomes
\begin{equation}
    U_{\mathrm{fb}}^{\mathrm S}(t)
    =\exp\!\left[-i\nu\bigl(S_x\cos\Delta t+S_y\sin\Delta t\bigr)\right].
    \label{eq_cic_rotating_x_feedback}
\end{equation}
Thus the feedback-pulse axis follows the coherent precession at angular frequency $\Delta$. The resulting Lindblad equation is
\begin{align}
    \dot\rho_{\mathrm S}(t)
    &=-i[\Delta S_z,\rho_{\mathrm S}(t)]+\mathcal D[J_{\mathrm S}(t)]\rho_{\mathrm S}(t)
    \nonumber\\
    &=-i[\Delta S_z,\rho_{\mathrm S}(t)]
    +\gamma\left(U_{\mathrm{fb}}^{\mathrm S}(t)S_-\rho_{\mathrm S}(t)S_+\bigl(U_{\mathrm{fb}}^{\mathrm S}(t)\bigr)^\dagger
    -\frac12\{S_+S_-,\rho_{\mathrm S}(t)\}\right).
    \label{eq_cic_feedback_master}
\end{align}
Here, the effective jump operator $J_{\mathrm S}(t)$
depends explicitly on time, so the CIC condition
discussed above for time-independent jump operators
does not apply directly. We therefore extend the
CIC condition to time-dependent jump operators
in the following subsection.

\subsection{Time-dependent CIC condition}
\label{subsec:cic_time_dependent}

For a time-independent Hamiltonian $H$ and a time-dependent jump operator $J_{\mathrm S}(t)$, we require the interaction-picture jump operator to vary only through an overall phase:
\begin{equation}
    J_{\mathrm I}(t)\equiv V^\dagger(t)J_{\mathrm S}(t)V(t)
    =e^{-i\omega t}J_{\mathrm I}(0),\qquad \omega\in\mathbb R,
    \label{eq_cic_covariance_condition}
\end{equation}
where $\omega$ is constant and $J_{\mathrm I}(0)=J_{\mathrm S}(0)$. For the feedback protocol above, $J_{\mathrm I}(0)=U_{\mathrm{fb}}^{\mathrm I}L$ is the fixed operator denoted by $J_{\mathrm I}$ in Eq.~\eqref{eq_cic_feedback_interaction}; $J_{\mathrm I}(t)$ additionally includes the time-dependent phase. Differentiating the transformed operator gives
\begin{align}
    \partial_tJ_{\mathrm I}(t)
    &=V^\dagger(t)\bigl(\partial_tJ_{\mathrm S}(t)+i[H,J_{\mathrm S}(t)]\bigr)V(t)
    \nonumber\\
    &=V^\dagger(t)\bigl(\partial_tJ_{\mathrm S}(t)-i[J_{\mathrm S}(t),H]\bigr)V(t).
    \label{eq_cic_rotated_derivative}
\end{align}
The phase dependence in
Eq.~\eqref{eq_cic_covariance_condition} requires
$\partial_tJ_{\mathrm I}(t)=-i\omega J_{\mathrm I}(t)$.
Combining this relation with
Eq.~\eqref{eq_cic_rotated_derivative} gives
\begin{equation}
    \partial_tJ_{\mathrm S}(t)-i[J_{\mathrm S}(t),H]=-i\omega J_{\mathrm S}(t),
\end{equation}
which, upon multiplication by $i$, yields
\begin{equation}
    i\partial_tJ_{\mathrm S}(t)+[J_{\mathrm S}(t),H]=\omega J_{\mathrm S}(t).
    \label{eq_cic_differential_condition}
\end{equation}
Equation~\eqref{eq_cic_differential_condition} is the CIC
condition for a time-dependent jump operator.
Conversely, this condition implies Eq.~\eqref{eq_cic_covariance_condition}. When $J_{\mathrm S}$ is time independent, it reduces to the original CIC condition $[J_{\mathrm S},H]=\omega J_{\mathrm S}$.

Since the overall phase cancels in the dissipator,
$\mathcal D[J_{\mathrm I}(t)]
=\mathcal D[J_{\mathrm I}(0)]$.
The interaction-picture master equation therefore becomes
$\dot\rho_{\mathrm I}
=\mathcal D[J_{\mathrm I}(0)]\rho_{\mathrm I}$.
Thus each jump contributes only a scalar phase to the evolution. As in the time-independent case, the jump statistics therefore coincide with those of the $H=0$ system with the fixed jump operator $J_{\mathrm I}(0)$.

\subsection{Application to our feedback model and the KUR}
\label{subsec:cic_statistics_kur}

For the protocol in Eq.~\eqref{eq_cic_rotating_feedback},
$\partial_tU_{\mathrm{fb}}^{\mathrm S}(t)=-i[H,U_{\mathrm{fb}}^{\mathrm S}(t)]$.
Using $J_{\mathrm S}(t)=U_{\mathrm{fb}}^{\mathrm S}(t)L$ and $[L,H]=\Delta L$, we obtain
\begin{align}
    i\partial_tJ_{\mathrm S}(t)&=[H,U_{\mathrm{fb}}^{\mathrm S}(t)]L,
    \nonumber\\
    [J_{\mathrm S}(t),H]
    &=U_{\mathrm{fb}}^{\mathrm S}(t)[L,H]+[U_{\mathrm{fb}}^{\mathrm S}(t),H]L
    \nonumber\\
    &=\Delta J_{\mathrm S}(t)-[H,U_{\mathrm{fb}}^{\mathrm S}(t)]L.
    \label{eq_cic_feedback_cancellation}
\end{align}
The commutator terms cancel in their sum, giving
\begin{equation}
    i\partial_tJ_{\mathrm S}(t)+[J_{\mathrm S}(t),H]
    =\Delta J_{\mathrm S}(t).
    \label{eq_cic_feedback_differential}
\end{equation}
Hence the time-dependent CIC condition Eq.~\eqref{eq_cic_differential_condition} holds with $\omega=\Delta$.

The CIC therefore ensures that the mean and variance
of $N_{\mathrm J}(\tau)$ coincide with those of the
$H=0$ dynamics in the main text:
\begin{equation}
    \langle N_{\mathrm J}(\tau)\rangle_{\mathrm S}
    =\langle N_{\mathrm J}(\tau)\rangle_{\mathrm I},
    \qquad
    \operatorname{Var}_{\mathrm S}[N_{\mathrm J}(\tau)]
    =\operatorname{Var}_{\mathrm I}[N_{\mathrm J}(\tau)].
\end{equation}
The dynamical activity also coincides in the two pictures:
\begin{equation}
    A_{\mathrm D}^{\mathrm S}(\tau)
    =\int_0^\tau dt\,\operatorname{Tr}[J_{\mathrm S}^\dagger(t)J_{\mathrm S}(t)\rho_{\mathrm S}(t)]
    =\gamma\int_0^\tau dt\,\langle S_+S_-\rangle_{\mathrm I}
    =A_{\mathrm D}^{\mathrm I}(\tau).
    \label{eq_cic_feedback_activity}
\end{equation}
So the End Matter KUR gives
\begin{equation}
    \frac{\operatorname{Var}_{\mathrm S}[N_{\mathrm J}(\tau)]}
    {[\tau\partial_\tau\langle N_{\mathrm J}(\tau)\rangle_{\mathrm S}]^2}
    \geq\frac{1}{A_{\mathrm D}^{\mathrm S}(\tau)}.
    \label{eq_cic_feedback_kur}
\end{equation}
Thus the counting precision, activity, and their scaling with $N$ are unchanged by retaining $H$ with the consistently transformed feedback protocol.

\section{Dynamical activity under independent emission}
\label{sec:independent_emission}

In the main text, collective emission described by the jump operator $L=\sqrt{\gamma}S_-$, with $S_-=\sum_{i=1}^N\sigma_-^{(i)}$, enables an $O(N^2)$ dynamical activity and Heisenberg scaling of the counting precision. Here we show that independent emission, described by separate jump operators proportional to $\sigma_-^{(i)}$, has a dynamical activity of at most $O(N)$, limiting the scaling of the counting fluctuation in the KUR to $O(N^{-1})$ at best.

Consider $N$ identical two-level systems, each emitting independently at rate $\gamma$. In the $H=0$ interaction picture used in the main text, the jump operators are $L_i=\sqrt{\gamma}\,\sigma_-^{(i)}$, and the Lindblad equation is
\begin{equation}
    \dot\rho
    =\sum_{i=1}^N\mathcal D[L_i]\rho
    =\gamma\sum_{i=1}^N
    \left(\sigma_-^{(i)}\rho\sigma_+^{(i)}
    -\frac12\{\sigma_+^{(i)}\sigma_-^{(i)},\rho\}\right).
    \label{eq_independent_master}
\end{equation}
The dynamical activity is the expected total number of jumps from all channels,
\begin{equation}
    A_{\mathrm{ind}}(\tau)
    =\int_0^\tau dt\,\sum_{i=1}^N
      \operatorname{Tr}[L_i^\dagger L_i\rho(t)]
    =\gamma\int_0^\tau dt\,\sum_{i=1}^N
      \langle\sigma_+^{(i)}\sigma_-^{(i)}\rangle.
    \label{eq_independent_activity}
\end{equation}
Since $\sigma_+^{(i)}\sigma_-^{(i)}$ is the excited-state projector of spin $i$, its expectation value lies between zero and one. Thus
\begin{equation}
    \frac{dA_{\mathrm{ind}}(t)}{dt}\leq\gamma N,
    \qquad
    A_{\mathrm{ind}}(\tau)\leq\gamma N\tau=O(N).
    \label{eq_independent_activity_bound}
\end{equation}
Unitary feedback after a jump replaces $L_i$ by $J_i=U_{\mathrm{fb},i}L_i$, but leaves $J_i^\dagger J_i=L_i^\dagger L_i$ unchanged. Therefore, even feedback acting on the ensemble cannot overcome Eq.~\eqref{eq_independent_activity_bound}.

Since the same form of the KUR holds for independent emission, with $A_{\mathrm D}(\tau)$ replaced by $A_{\mathrm{ind}}(\tau)$, the upper bound on $A_{\mathrm{ind}}(\tau)$ excludes the $O(N^{-2})$ scaling of the counting fluctuation. Thus, independent emission cannot achieve Heisenberg
scaling, and collective emission is essential to
achieving this scaling.

\section{Derivation of the mean-field equations}
\label{sec:mf_derivation}

We derive the mean-field equations in Eqs.~\refMainMFEnoFB and \refMainMFEwithFB of the main text. While the main text focuses on $m_z$, which determines the dynamical activity, here we also derive the equations for $m_x$ and $m_y$. For simplicity of notation, we work in the interaction picture with respect to $H=\Delta S_z$, so the Hamiltonian term vanishes. We carry out the derivation with feedback; the dynamics without feedback is recovered by setting $\nu=0$.

The feedback dynamics is
\begin{equation}
    \dot\rho=\gamma\left(U_{\mathrm{fb}}S_-\rho S_+U_{\mathrm{fb}}^\dagger
    -\frac12\{S_+S_-,\rho\}\right),
    \qquad U_{\mathrm{fb}}=e^{-i\nu F},\qquad
    F=\sin\phi_0\,S_x-\cos\phi_0\,S_y.
    \label{eq_sm_feedback_master}
\end{equation}
For an arbitrary system operator $O$, its expectation evolves as
$d\langle O\rangle/dt=\langle\mathcal L_{\mathrm{fb}}^\dagger[O]\rangle$, where
\begin{equation}
    \mathcal L_{\mathrm{fb}}^\dagger[O]
    =\gamma\left(S_+U_{\mathrm{fb}}^\dagger OU_{\mathrm{fb}}S_-
    -\frac12\{S_+S_-,O\}\right).
    \label{eq_sm_adjoint_liouvillian}
\end{equation}
Adding and subtracting $\gamma S_+OS_-$ separates the collective decay and feedback contributions:
\begin{equation}
    \mathcal L_{\mathrm{fb}}^\dagger[O]
    =\mathcal L_0^\dagger[O]
    +\gamma S_+(U_{\mathrm{fb}}^\dagger OU_{\mathrm{fb}}-O)S_-,
    \qquad
    \mathcal L_0^\dagger[O]
    =\frac{\gamma}{2}\left(S_+[O,S_-]+[S_+,O]S_-\right).
    \label{eq_sm_split_adjoint}
\end{equation}

\subsection{Operator dynamics under feedback}
\label{subsec:feedback_mf_part}

Using $[S_x,S_-]=S_z$, $[S_+,S_x]=S_z$, $[S_y,S_-]=-iS_z$, $[S_+,S_y]=iS_z$, $[S_z,S_-]=-S_-$, and $[S_+,S_z]=-S_+$, we obtain
\begin{align}
    \mathcal L_0^\dagger[S_x]
    &=\frac{\gamma}{2}\left(\{S_x,S_z\}-S_x\right),
    \nonumber\\
    \mathcal L_0^\dagger[S_y]
    &=\frac{\gamma}{2}\left(\{S_y,S_z\}-S_y\right),
    \nonumber\\
    \mathcal L_0^\dagger[S_z]
    &=-\gamma S_+S_-
    =-\gamma(S_x^2+S_y^2+S_z).
    \label{eq_sm_exact_collective_decay}
\end{align}
We define
\begin{equation}
    S_{\parallel}=\cos\phi_0\,S_x+\sin\phi_0\,S_y.
    \label{eq_sm_parallel_spin}
\end{equation}
The commutators $[F,S_{\parallel}]=iS_z$ and $[F,S_z]=-iS_{\parallel}$ give
$\partial_\nu(U_{\mathrm{fb}}^\dagger S_{\parallel}U_{\mathrm{fb}})=-U_{\mathrm{fb}}^\dagger S_zU_{\mathrm{fb}}$ and
$\partial_\nu(U_{\mathrm{fb}}^\dagger S_zU_{\mathrm{fb}})=U_{\mathrm{fb}}^\dagger S_{\parallel}U_{\mathrm{fb}}$.
Solving these differential equations with the initial conditions
$U_{\mathrm{fb}}^\dagger S_{\parallel}U_{\mathrm{fb}}=S_{\parallel}$
and $U_{\mathrm{fb}}^\dagger S_zU_{\mathrm{fb}}=S_z$ at $\nu=0$,
and noting that $[F,U_{\mathrm{fb}}]=0$, we obtain
\begin{equation}
    U_{\mathrm{fb}}^\dagger S_{\parallel}U_{\mathrm{fb}}=S_{\parallel}\cos\nu-S_z\sin\nu,
    \qquad U_{\mathrm{fb}}^\dagger S_zU_{\mathrm{fb}}=S_{\parallel}\sin\nu+S_z\cos\nu,
    \qquad U_{\mathrm{fb}}^\dagger FU_{\mathrm{fb}}=F.
    \label{eq_sm_rotation}
\end{equation}
Since $S_x=\cos\phi_0\,S_{\parallel}+\sin\phi_0\,F$ and
$S_y=\sin\phi_0\,S_{\parallel}-\cos\phi_0\,F$, the changes in all three components are
\begin{align}
    U_{\mathrm{fb}}^\dagger S_xU_{\mathrm{fb}}-S_x
    &=\cos\phi_0\bigl[(\cos\nu-1)S_{\parallel}-\sin\nu\,S_z\bigr],
    \nonumber\\
    U_{\mathrm{fb}}^\dagger S_yU_{\mathrm{fb}}-S_y
    &=\sin\phi_0\bigl[(\cos\nu-1)S_{\parallel}-\sin\nu\,S_z\bigr],
    \nonumber\\
    U_{\mathrm{fb}}^\dagger S_zU_{\mathrm{fb}}-S_z
    &=\sin\nu\,S_{\parallel}+(\cos\nu-1)S_z.
    \label{eq_sm_kick_operator_differences}
\end{align}

\subsection{Mean-field equations for the magnetization}
\label{subsec:mf_factorization}

We introduce the normalized magnetization
\begin{equation}
    m_\alpha=\frac{\langle S_\alpha\rangle}{S},
    \qquad S=\frac{N}{2},
    \qquad \alpha=x,y,z.
\end{equation}
The exact equations derived above contain quadratic and cubic
spin moments, so the evolution of the mean magnetization depends
on higher-order moments. To obtain equations involving only
$m_x$, $m_y$, and $m_z$, we apply the mean-field approximation:
after evaluating the commutators exactly, we neglect fluctuation
contributions to the remaining moments and factorize them into
products of mean values.

For the quadratic moments in the collective-decay contribution,
this gives
\begin{equation}
    \langle\{S_\alpha,S_z\}\rangle
    \simeq2\langle S_\alpha\rangle\langle S_z\rangle
    =2S^2m_\alpha m_z,
    \qquad
    \langle S_\alpha^2\rangle
    \simeq\langle S_\alpha\rangle^2
    =S^2m_\alpha^2,
    \qquad \alpha=x,y.
\end{equation}
Similarly, the cubic moments in the feedback contribution
are approximated as
\begin{equation}
    \langle S_+S_\beta S_-\rangle
    \simeq
    \langle S_+\rangle\langle S_\beta\rangle
    \langle S_-\rangle,
    \qquad \beta=x,y,z.
    \label{eq_sm_three_body_factorization}
\end{equation}
Since $S_\pm=S_x\pm iS_y$, we have
$\langle S_\pm\rangle=S(m_x\pm im_y)$, and hence
\begin{equation}
    \langle S_+S_\beta S_-\rangle
    \simeq S^3(m_x+im_y)m_\beta(m_x-im_y)
    =S^3(m_x^2+m_y^2)m_\beta.
\end{equation}
For compactness, we define
\begin{equation}
    q=m_x^2+m_y^2,
    \qquad
    m_{\parallel}=\cos\phi_0\,m_x+\sin\phi_0\,m_y,
    \label{eq_sm_magnetization}
\end{equation}
so that
$\langle S_+S_\beta S_-\rangle\simeq S^3q\,m_\beta$.
Combining these approximations with
Eqs.~\eqref{eq_sm_exact_collective_decay} and
\eqref{eq_sm_kick_operator_differences}, and dividing
$d\langle S_\alpha\rangle/dt$ by $S=N/2$, we obtain
\begin{align}
    \dot m_x
    &\simeq\frac{\gamma N}{2}m_xm_z-\frac{\gamma}{2}m_x
    +\frac{\gamma N^2}{4}q\cos\phi_0
    \bigl[(\cos\nu-1)m_{\parallel}-m_z\sin\nu\bigr],
    \label{eq_sm_mf_full_x}\\
    \dot m_y
    &\simeq\frac{\gamma N}{2}m_ym_z-\frac{\gamma}{2}m_y
    +\frac{\gamma N^2}{4}q\sin\phi_0
    \bigl[(\cos\nu-1)m_{\parallel}-m_z\sin\nu\bigr],
    \label{eq_sm_mf_full_y}\\
    \dot m_z
    &\simeq-\frac{\gamma N}{2}q-\gamma m_z
    +\frac{\gamma N^2}{4}q
    \bigl[m_{\parallel}\sin\nu+m_z(\cos\nu-1)\bigr].
    \label{eq_sm_mf_full_z}
\end{align}
The equation for $m_z$, Eq.~\eqref{eq_sm_mf_full_z},
reproduces Eq.~\refMainMFEwithFB of the main text.

\subsection{Conservation of the spin length}
\label{subsec:mf_spin_length}

We next show that, for $\nu=c/N$ with $c$ independent of $N$, the leading-order mean-field dynamics preserves $m_x^2+m_y^2+m_z^2=1$ when this relation holds initially. Using $\sin\nu=c/N+O(N^{-3})$ and $\cos\nu-1=-c^2/(2N^2)+O(N^{-4})$, and retaining the leading $O(N)$ terms in Eqs.~\eqref{eq_sm_mf_full_x}--\eqref{eq_sm_mf_full_z}, we obtain
\begin{align}
    \dot m_x
    &=\frac{\gamma N}{2}\left(m_xm_z-\frac{c}{2}q\,m_z\cos\phi_0\right),
    \nonumber\\
    \dot m_y
    &=\frac{\gamma N}{2}\left(m_ym_z-\frac{c}{2}q\,m_z\sin\phi_0\right),
    \nonumber\\
    \dot m_z
    &=\frac{\gamma N}{2}q\left(\frac{c}{2}m_{\parallel}-1\right).
    \label{eq_sm_feedback_leading_mf}
\end{align}
These equations give
\begin{align}
    \frac{d}{dt}(m_x^2+m_y^2+m_z^2)
    &=2m_x\dot m_x+2m_y\dot m_y+2m_z\dot m_z
    \nonumber\\
    &=\gamma N m_z(m_x^2+m_y^2-q)
    +\frac{\gamma cN}{2}q\,m_z(m_{\parallel}-m_x\cos\phi_0-m_y\sin\phi_0)
    =0,
    \label{eq_sm_spin_length_cancellation}
\end{align}
since $q=m_x^2+m_y^2$ and $m_{\parallel}=m_x\cos\phi_0+m_y\sin\phi_0$. Thus, for an initial state with unit normalized spin length,
\begin{equation}
    m_x^2+m_y^2+m_z^2=1
    \label{eq_sm_leading_spin_length}
\end{equation}
holds throughout the leading-order mean-field evolution, allowing us to use $q=1-m_z^2$.

In particular, $c=0$ corresponds to dynamics without feedback, for which
\begin{equation}
    \dot m_z\simeq-\frac{\gamma N}{2}q
    =-\frac{\gamma N}{2}(1-m_z^2),
    \label{eq_sm_uncontrolled_mf}
\end{equation}
recovering Eq.~\refMainMFEnoFB of the main text.

\section{Mean-field evaluation of the dynamical activity and numerical accuracy}
\label{sec:activity_validation}

\subsection{Mean-field expression for the activity}
\label{subsec:activity_derivation}

We derive the mean-field expression for $A_{\mathrm D}(\tau)$ in Eq.~\refMainMFDA of the main text. The unitary kick leaves the jump-rate operator unchanged, $(U_{\mathrm{fb}}L)^\dagger(U_{\mathrm{fb}}L)=L^\dagger L$, so both with and without feedback we have
\begin{equation}
    A_{\mathrm D}(\tau)
    =\gamma\int_0^\tau dt\,\langle S_+S_-\rangle.
    \label{eq_sm_activity_def}
\end{equation}
Equation~\eqref{eq_sm_exact_collective_decay} gives
$\langle S_+S_-\rangle=\langle S_x^2\rangle+\langle S_y^2\rangle+\langle S_z\rangle$.
Since $\langle S_z\rangle=Nm_z/2$, it remains to evaluate
$\langle S_x^2\rangle$ and $\langle S_y^2\rangle$.

For either $\alpha=x$ or $y$, expanding $S_\alpha=\frac12\sum_{i=1}^N\sigma_\alpha^{(i)}$ gives
\begin{align}
    \langle S_\alpha^2\rangle
    &=\frac14\sum_{i=1}^N
      \left\langle(\sigma_\alpha^{(i)})^2\right\rangle
      +\frac14\sum_{i\ne j}
      \left\langle\sigma_\alpha^{(i)}\sigma_\alpha^{(j)}\right\rangle
    \nonumber\\
    &=\frac N4+\frac14\sum_{i\ne j}
      \left\langle\sigma_\alpha^{(i)}\sigma_\alpha^{(j)}\right\rangle.
    \label{eq_sm_activity_site_decomposition}
\end{align}
For distinct indices, we use the mean-field approximation. Permutation symmetry first gives
\begin{equation}
    \left\langle\sigma_\alpha^{(i)}\right\rangle
    =\frac{2}{N}\langle S_\alpha\rangle=m_\alpha
    \label{eq_sm_activity_single_spin_mean}
\end{equation}
at every site, so factorizing the two-spin expectation yields
\begin{equation}
    \left\langle\sigma_\alpha^{(i)}\sigma_\alpha^{(j)}\right\rangle
    \simeq\left\langle\sigma_\alpha^{(i)}\right\rangle
    \left\langle\sigma_\alpha^{(j)}\right\rangle
    =m_\alpha^2,\qquad i\ne j.
    \label{eq_sm_activity_pair_factorization}
\end{equation}
The sum contains $N(N-1)$ such ordered pairs. Substitution into Eq.~\eqref{eq_sm_activity_site_decomposition} then gives
\begin{equation}
    \langle S_\alpha^2\rangle
    \simeq\frac N4+\frac{N(N-1)}4m_\alpha^2
    =\frac{N^2}{4}m_\alpha^2+\frac N4(1-m_\alpha^2).
    \label{eq_sm_activity_second_moment_mf}
\end{equation}

Applying this result to both transverse components and adding $\langle S_z\rangle=Nm_z/2$, we obtain
\begin{align}
    \langle S_+S_-\rangle
    &\simeq\frac{N^2}{4}(m_x^2+m_y^2)
    +\frac N4(2-m_x^2-m_y^2)+\frac N2m_z
    \nonumber\\
    &=\frac{N^2}{4}q+\frac N4(2-q+2m_z),
    \qquad q=m_x^2+m_y^2.
    \label{eq_sm_activity_rate_order}
\end{align}
Retaining only the leading $O(N^2)$ term, we obtain
$\langle S_+S_-\rangle\simeq N^2q/4$.

For initial states with unit normalized spin length, Sec.~\ref{subsec:mf_spin_length} shows that $q=1-m_z^2$ is preserved by the leading-order dynamics for $\nu=c/N$, including the case without feedback ($c=0$). Therefore,
\begin{equation}
    A_{\mathrm D}(\tau)
    \simeq\frac{\gamma N^2}{4}\int_0^\tau dt\,q(t)
    =\frac{\gamma N^2}{4}\int_0^\tau dt\,[1-m_z(t)^2],
    \label{eq_sm_activity_mf}
\end{equation}
which is Eq.~\refMainMFDA of the main text.

\subsection{Comparison with the master-equation calculation}
\label{subsec:activity_numerical_accuracy}

To assess the accuracy of Eq.~\eqref{eq_sm_activity_mf},
we numerically solve the master equation and compare
$A_{\mathrm D}(\tau)$ from Eq.~\eqref{eq_sm_activity_def}
with the following approximation:
\begin{equation}
    \widetilde A_{\mathrm D}(\tau)
    \equiv\frac{\gamma N^2}{4}\int_0^\tau dt\,[1-m_z(t)^2].
    \label{eq_sm_activity_estimate}
\end{equation}
Here the tilde identifies this approximation to the exact activity.

\begin{figure}[t]
    \centering
    \includegraphics[width=0.88\textwidth]{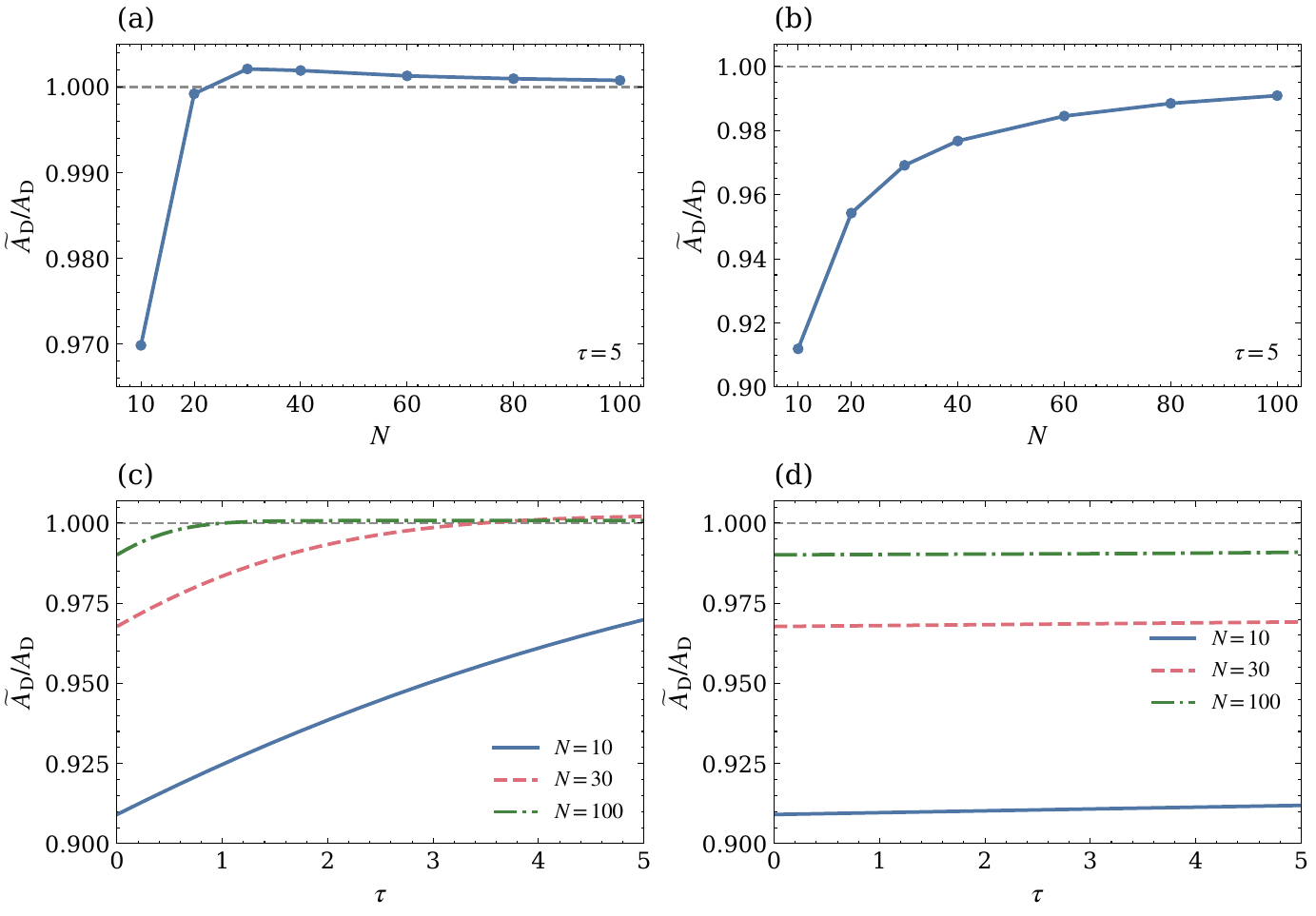}
    \caption{Accuracy of the magnetization-based activity approximation. Panels (a) and (b) show $\widetilde A_{\mathrm D}(\tau=5)/A_{\mathrm D}(\tau=5)$ versus $N$ without and with feedback, respectively. Panels (c) and (d) show the same ratio as a function of observation time for $N=10,30,100$ without and with feedback, respectively. The dashed horizontal lines indicate unity.  The parameters are $\gamma=0.02$, $\Delta=0$, and the initial magnetization $(0,1,0)$; $\nu=0$ without feedback and $\nu=\arcsin(2/N)$ with feedback.}
    \label{fig_sm_activity_accuracy}
\end{figure}

Figure~\ref{fig_sm_activity_accuracy} compares
$\widetilde A_{\mathrm D}(\tau)/A_{\mathrm D}(\tau)$ without feedback
[panels (a) and (c)] and with feedback [panels (b) and (d)].
The parameters match the main-text simulations:
$\gamma=0.02$, $\Delta=0$, and the spin-coherent initial state
with $(m_x,m_y,m_z)=(0,1,0)$.
We use $\nu=0$ without feedback and $\nu=\arcsin(2/N)$
with feedback.
Panels (a) and (b) show the $N$ dependence at $\tau=5$
for $N=10$, $20$, $30$, $40$, $60$, $80$, and $100$.
Without feedback [panel (a)], the relative deviation
$|\widetilde A_{\mathrm D}/A_{\mathrm D}-1|$
is approximately $3.0\%$ for $N=10$ and below $0.22\%$
for all displayed $N\geq20$.
With feedback [panel (b)], it decreases from approximately
$8.8\%$ at $N=10$ to $0.91\%$ at $N=100$.
Panels (c) and (d) show the time dependence for
$N=10,30,100$, without and with feedback, respectively.
Over all sampled times $0<\tau\leq5$, the deviation is
below $9.1\%$ for all tested sizes and below $1.0\%$
for $N=100$ in both cases.
These results quantify the accuracy of the activity
approximation over the parameter range studied here.

%